\documentclass[12pt]{article}
\usepackage{epsfig}
\usepackage{amssymb}

\def\beq{\begin{equation}}
\def\eeq#1{\label{#1}\end{equation}}
\def\eeqn{\end{equation}}

\def\beqa{\begin{eqnarray}}
\def\eeqa#1{\label{#1}\end{eqnarray}}
\def\eeqan{\end{eqnarray}}

\let\bar=\overbar

\def\vev#1{\langle #1 \rangle}

\def\Dslash{\not{\hbox{\kern-4pt $D$}}}
\def\dslash{\not{\hbox{\kern-2pt $\del$}}}

\def\msb{{\bar{\ssstyle M \kern -1pt S}}}

\def\Title#1{\begin{center} {\Large {\bf #1} } \end{center}}

\newcommand{\eq}[1]{(\ref{#1})}

\newcommand{\lr}[1]{ \left( #1 \right) }

\newcommand{\rvac}{ \, | 0 \rangle }
\newcommand{\lvac}{ \langle 0 | \, }

\newcommand{\logo}{\\ \vskip -24mm
\leftline{\includegraphics[scale=0.3,clip=false]{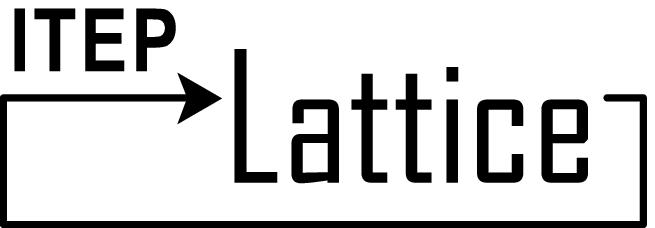}} \vskip 16mm}

\newcommand{\prep}{~\\ \vskip -33mm\rightline{\normalsize{ITEP-LAT/2009-08}} \vskip 20mm}

\begin{document}

\renewcommand{\thefootnote}{\fnsymbol{footnote}}

\Title{Lattice QCD in strong magnetic fields\footnote{Talk given by
M.N.~Chernodub at the 10th Workshop on Non-Perturbative QCD, June 8-12, 2009, Paris, France.}\logo\prep}

\setcounter{footnote}{3}

\bigskip\bigskip


\begin{raggedright}

{\it P.V.Buividovich$^{\,ab}$, M.N.Chernodub$^{\,cdb}$, E.V.Luschevskaya$^{\,b}$, M.I.Polikarpov$^{\,b}$\\
${}^a$ JIPNR ``Sosny'', National Academy of Science, Krasin str. 99, Minsk, Belarus\\
${}^b$ ITEP, B. Cheremushkinskaya 25, Moscow, 117218 Russia\\
${}^c$ CNRS, LMPT, F\'ed\'eration Denis Poisson, Universit\'e de Tours, 37200 France\\
${}^d$ DMPA, University of Gent, Krijgslaan 281, S9, B-9000 Gent, Belgium
}
\bigskip\bigskip
\end{raggedright}

\renewcommand{\thefootnote}{\arabic{footnote}}
\setcounter{footnote}{0}

\abstract{Vacuum of Quantum Chromodynamics in very strong (hadron-scale) magnetic fields exhibits many interesting nonperturbative effects.
Some of these effects can be studied with the help of lattice simulations in quenched QCD. We review
our recent results demonstrating that very strong external magnetic fields lead to
(1) the enhancement of the chiral symmetry breaking [the quark condensate rises with the increase of the external magnetic field];
(2) the chiral magnetization of the QCD vacuum [spins of the quarks turn parallel to the external~field];
(3) the chiral magnetic effect [a CP-odd generation of the electric current of quarks directed along the magnetic field];
(4) a CP-odd generation of the anomalous quark electric dipole moment along the axis of magnetic field.
The first three effects were already predicted theoretically, and subsequently observed numerically in our simulations,
while the fourth effect is a new result.}

\section{Introduction}
\subsection*{Motivation}

Noncentral heavy-ion collisions may create very strong magnetic field due to relative motion of electrically
charged ions and the products of the collision. This fact ignited the interest of the scientific community towards
an investigation of the properties of the strongly interacting matter exposed to the magnetic field. The QCD effects
should be visible because the strength of the created magnetic field may be of the order of the hadron-scale, or higher.
For example, at first moments ($\tau \sim 1$\,fm/c) of Au-Au collision at the Relativistic Heavy Ion Collider (RHIC)
the strength of the magnetic field may reach $e B \sim (10 - 100 \, \mbox{MeV})^2$~\cite{Kharzeev:08:1,Skokov}.
The strong magnetic fields will also be created in the future experiments at the Facility for Antiproton and
Ion Research (FAIR) at GSI, at the ALICE experiment at LHC, and at the Nuclotron-based Ion Collider fAcility (NICK) at Dubna.

\subsection*{Vacuum effects due to strong magnetic field in QCD}

There are various vacuum effects which appear due to the strong external magnetic fields (here we do not discuss
interesting phenomena that may emerge in a dense quark matter).

\vskip 1mm

{\bf Enhancement and shift of the chiral phase transition.} Theoretically, the very strong
magnetic fields may significantly modify the QCD phase diagram~\cite{Agasian,Fraga:08:1:2:3}. The external magnetic field
increases the transition temperature and enhances the strength of the (phase) transition from the chirally broken
(low temperature) phase to the chirally restored (high temperature) phase. Moreover, in the external magnetic field
the transition -- which is a smooth crossover in the absence of the fields -- becomes a first order
transition~\cite{Agasian,Fraga:08:1:2:3}.

\vskip 1mm

{\bf Enhancement of the chiral symmetry breaking.} The magnetic
field stabilizes the chirally broken (low-temperature) phase of QCD by
enlarging the value of the chiral condensate and enhancing the
chiral symmetry breaking. This result was obtained in various
approaches using the chiral perturbation
theory~\cite{Smilga:97:1,Agasian:1999sx,Cohen:2007bt}, the
Nambu-Jona-Lasinio
model~\cite{Gusynin:1995nb}, the linear
sigma model~\cite{Goyal:1999ye} and the AdS/QCD dual
description~\cite{Zayakin:08:1}. In lattice simulations this effect
was observed in our recent work~\cite{ref:condensate}. In
Section~\ref{sec:chiral} we discuss main results
of~\cite{ref:condensate}.

\vskip 1mm

{\bf Chiral magnetization.} Another effect of the strong enough magnetic field
is a (para)magnetic response of the QCD vacuum: the
external fields polarize the spins (or, equivalently, the magnetic moments) of the quarks and antiquarks,
leading to the chiral magnetization of the vacuum. At relatively low magnetic fields the chiral
magnetization is proportional to the strength of the magnetic field. The coefficient of the
proportionality, called the chiral magnetic susceptibility, was first discussed in Ref.~\cite{ref:IS} in order to analyze
phenomenologically interesting nucleon magnetic moments. The value of the magnetic susceptibility -- which was
estimated using various analytical approaches~\cite{ref:others:OPE,Dorokhov:alone,Dorokhov,ref:Vainshtein,ref:Sasha,ref:Kim} --
can be measured in experiments on lepton pair photoproduction~\cite{ref:phenomenology},
in radiative heavy meson decays~\cite{Rohrwild:2007yt}, etc. At higher magnetic fields the behavior of the
magnetization deviates from a linear function as it gets affected by logarithmic corrections at moderately
strong fields~\cite{ref:Thomas}. At asymptotically strong fields the magnetization reaches an upper bound.
The first lattice investigation of the chiral magnetization was performed in Ref.~\cite{ref:magnetization}.
The results of \cite{ref:magnetization} are briefly described in Section~\ref{sec:magnetization}.

\vskip 1mm

{\bf Chiral Magnetic Effect.} The magnetic fields may lead to quite unusual effects due to the nontrivial topological
structure of the QCD vacuum~\cite{Kharzeev:08:1,Kharzeev:98:1}.
We discuss a particular realization of the Chiral Magnetic Effect (CME), which gives rise to a
generation of an {\it electric} current along the direction of the {\it magnetic} field in
a nontrivial topological backgrounds  of gluons~\cite{Kharzeev:08:1}. The potentially observed feature
of the CME is a non-statistical asymmetry in the number of positively and negatively charged
particles emitted on different sides of the reaction plane in the heavy-ion collisions~\cite{Voloshin:04:1}.
There are preliminary indications that this $\mathcal{CP}$-odd effect has been indeed observed by
the STAR Collaboration in experiments at RHIC \cite{Voloshin:08:1}. In our lattice
studies we found the existence of the CME both at vacuum and thermal ground states of gluon fields,
and at specially prepared configurations with nontrivial topological charge~\cite{ref:CME}.
The main results of Ref.~\cite{ref:CME} are given in Section~\ref{sec:CME}.

\vskip 3mm

{\bf Quark electric dipole moment.} Another manifestation of a nontrivial topological
structure of the QCD vacuum is an appearance of an {\it electric} dipole moment of a quark along the direction
of the external {\it magnetic} field. This $\mathcal{CP}$-odd effect -- which is a spin analogue of the
Chiral Magnetic Effect -- is observed in lattice simulations in Ref.~\cite{ref:TBA}.
The lattice evidence of the generation of the quark's electric dipole moment is discussed in Section~\ref{sec:EDM}.

\subsection*{Physical setup and technical details of simulations}

We utilize lattice QCD with the simplest $SU(2)$ gauge group because all studied effects originate in the chiral sector of QCD
where the number of colors is not crucial. In our simulations only valence quarks interact with the
electromagnetic field, and the effects of the virtual quarks on gluons are neglected. Indeed, the inclusion of dynamical
(sea) quarks makes the simulations computationally difficult, while the essential features of all mentioned
effects are visible in the quenched limit as well (an extended discussion is given in \cite{ref:condensate,ref:magnetization,ref:CME,Luschevskaya:08:1}).

The chiral effects are best studied with massless fermions. In order to implement chirally symmetric massless fermions on the lattice,
we use Neuberger's overlap Dirac operator~\cite{Neuberger:98:1}. We reduce the ultraviolet lattice artifacts using the tadpole-improved
Symanzik action for the gluons. We introduce the magnetic field $B_\mu = B \delta_{\mu3}$ into the Dirac operator by
substituting $su\lr{2}$-valued vector potential $A_{\mu}$ with $u\lr{2}$-valued field. In infinite volume such substitution is
$A_{\mu}^{ij} \rightarrow A_{\mu}^{ij} + C_{\mu} \delta^{ij}$, where $C_{\mu} = B \lr{x_{2} \delta_{\mu1} - x_{1} \delta_{\mu2}}/2$
is the external $U(1)$ gauge field. In the finite volume $L^4$ with periodic boundary conditions an additional $x$-dependent boundary twist for the
fermion fields should be introduced~\cite{Wiese:08:1}, and the uniform magnetic field is forced to take
quantized values~\cite{Damgaard:88:1}, $q B = 2 \pi \, k/L^{2}$, where $k \in \mathbb{Z}$ and $q = 1/3\, |e|$
is the absolute value of the electric charge of the $d$-quark. Note that in all figures below we indicate the magnetic field
strength in units of $q B \equiv e B/3$  and not in $eB$. In our simulations the {\it lowest} magnetic
field takes values $\sqrt{e B} \sim (400 - 600 \, \mbox{MeV})^2$. Such fields are stronger than those expected at RHIC,
while they are of same order as the magnetic fields that will presumably be created at the LHC collisions~\cite{Skokov}.

In order to check the finite-volume effects we use two spatial lattice volumes, $14^3$ and $16^3$.
We study the system at three different temperatures,
$T/T_c=0, \, 0.82, \, 1.12$ (the critical temperature in $SU(2)$ gauge theory
is $T_c \approx 310$\,MeV). In order to make sure that the
influence of the finite ultraviolet cutoff is under proper control, we utilize
three lattice spacings, $a = 0.089\,\mbox{fm},\, 0.103\,\mbox{fm},\, 0.128\,\mbox{fm}$.
Our spatial volumes range from $L^3 = (1.4\,\mbox{fm})^3$ to $(2\,\mbox{fm})^3$.

\clearpage
\section{Enhancement of chiral condensate}
\label{sec:chiral}

The chiral condensate
\begin{eqnarray}
\Sigma \equiv - \lvac \bar{q} q \rvac\,,
\label{eq:Sigma}
\end{eqnarray}
is commonly used order parameter for the chiral symmetry breaking in the theory with massless quarks.
The condensate is zero if the chiral symmetry is unbroken. If the strength of magnetic field is
larger than the pion mass (which is zero for massless quarks) but is still much smaller
than the hadronic scale, one can use the chiral perturbation theory to calculate the field
dependence of the chiral condensate. According to the original work~\cite{Smilga:97:1},
in a leading order in $B$, the chiral condensate $\Sigma\lr{B}$ rises linearly with the field strength:
\begin{eqnarray}
\label{cc_vs_B_chPT}
\Sigma\lr{B} = \Sigma\lr{0}\, \lr{1 + \frac{e B \ln{2}}{16 \pi^{2} F_{\pi}^{2}} }\,,
\end{eqnarray}
where $F_{\pi} \approx 130 \, \mbox{MeV}$ is the pion decay constant. The corrections due to
nonvanishing pion mass may also be calculated~\cite{Cohen:2007bt}.

The condensate~(\ref{eq:Sigma}) can be calculated using the Banks-Casher formula \cite{Banks:80:1},
\begin{eqnarray}
\label{BanksCasher}
\Sigma = \lim \limits_{\lambda \rightarrow 0} \lim \limits_{V \rightarrow \infty} \, \frac{\pi \rho\lr{\lambda}}{V}
\end{eqnarray}
where $V$ is the total four-volume of Euclidean space-time. The density $\rho\lr{\lambda}$ of eigenvalues $\lambda_{n}$
of the Dirac operator $\mathcal{D} = \gamma^{\mu} \, \lr{\partial_{\mu} - i A_{\mu}}$ is defined by
\begin{eqnarray}
\mathcal{D} \psi_{n} = \lambda_{n} \psi_{n}\,,\qquad \rho\lr{\lambda} = \vev{ \sum \limits_{n} \delta\lr{\lambda - \lambda_{n}} }\,,
\end{eqnarray}
where $\psi_{n}$ are the eigenmodes of the Dirac operator.

\begin{figure}[htb]
\begin{center}
\begin{tabular}{cc}
\hskip -7mm \includegraphics[height=3.1in, angle=-90]{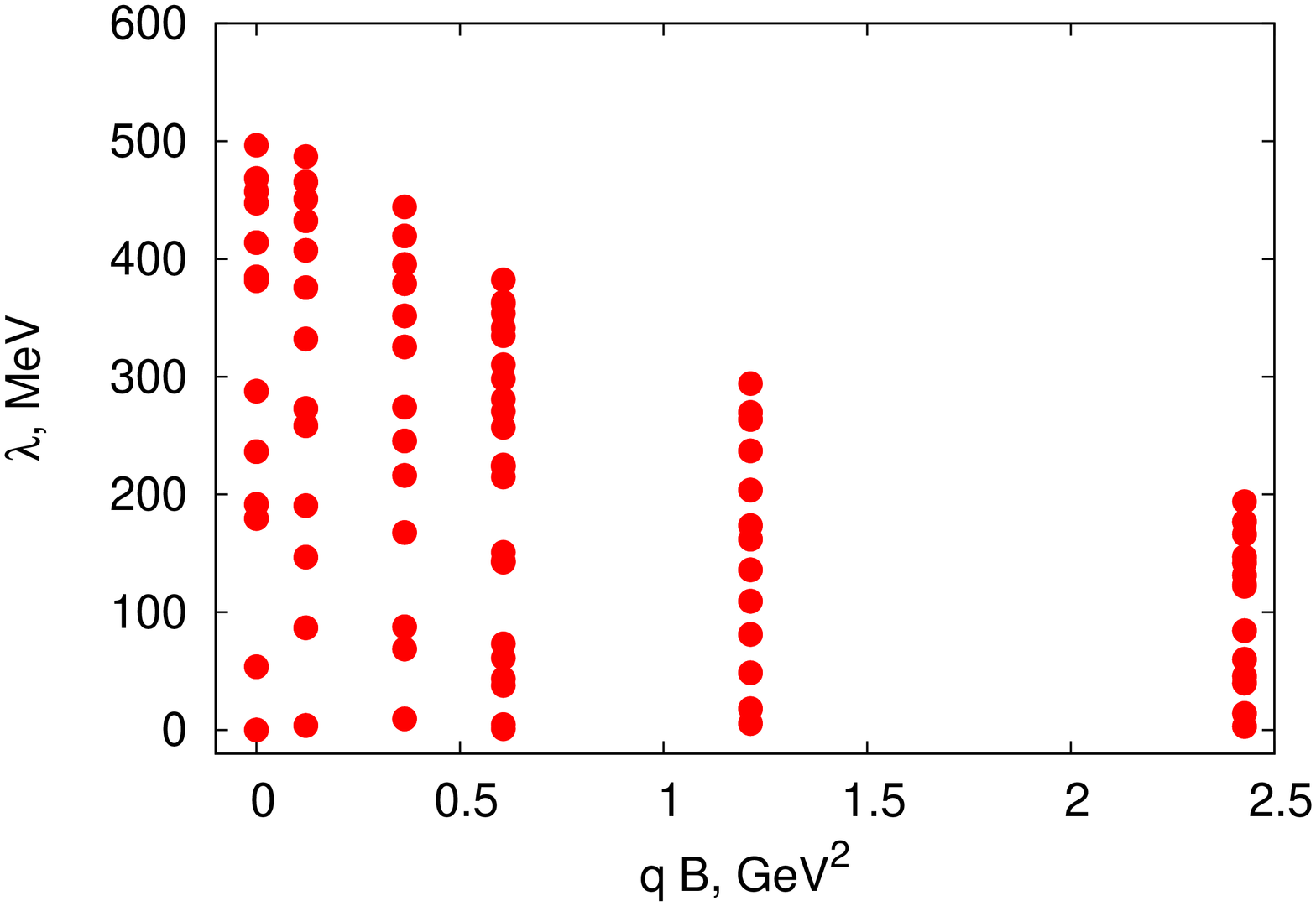} &
\hskip -5mm \includegraphics[height=3.1in, angle=-90]{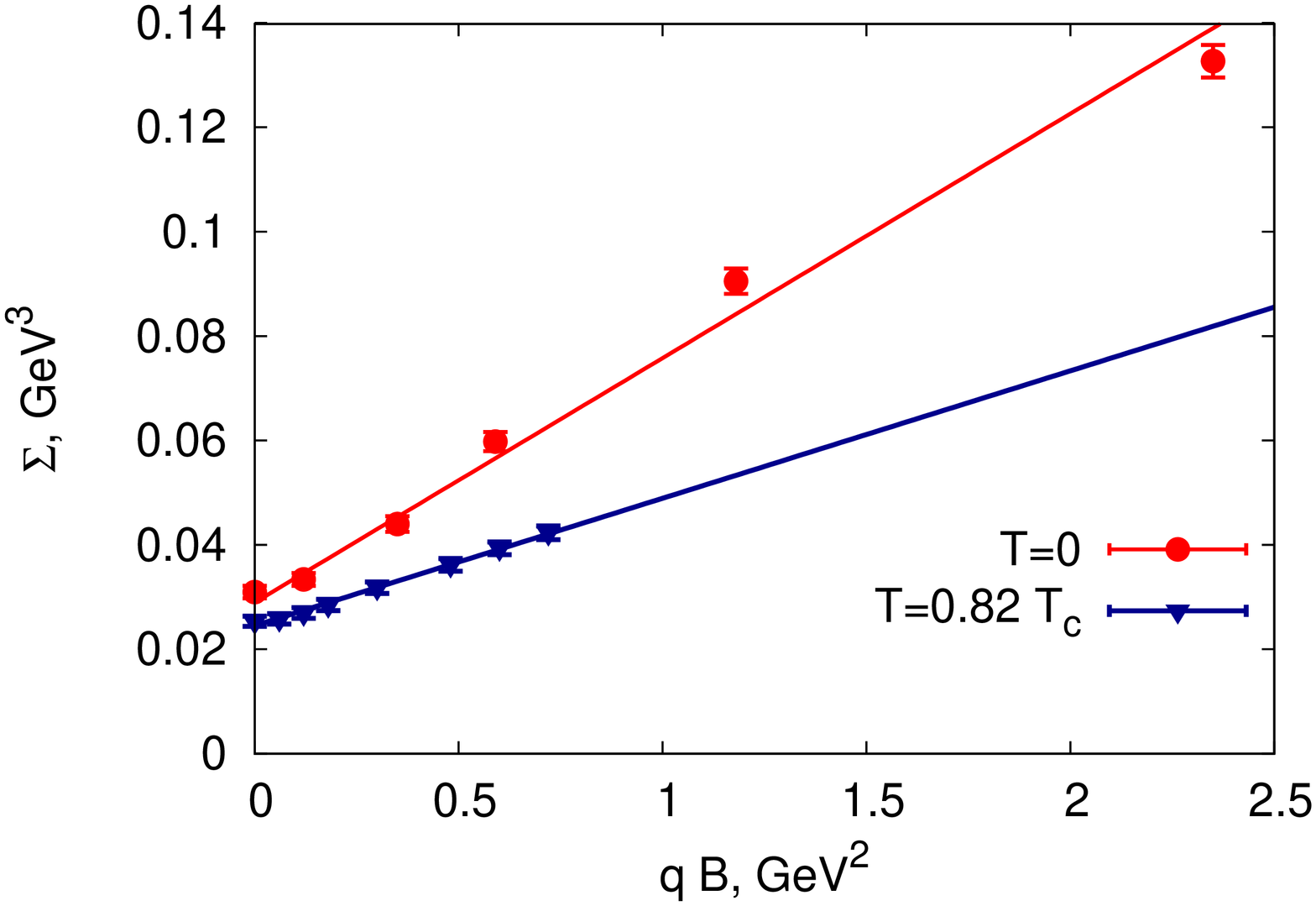}
\end{tabular}
\caption{(left) The spectrum of lowest twelve eigenvalues $\lambda$ (including the zero mode $\lambda=0$) of the Dirac operator in a background of
a typical gluon configuration vs the magnetic field $q B$.
(right) The chiral condensate vs $q B$ at two different tem\-pera\-tu\-res~$T$.
The solid lines are the linear fits by the function~(\ref{fit_fun}). Figures are from~\cite{ref:condensate}.}
\label{fig:condensate}
\end{center}
\end{figure}

According to Eq.~\eq{BanksCasher} the enhancement of the chiral condensate in the magnetic field means that the Dirac eigenvalues
in the vicinity of $\lambda=0$ should become denser as the magnetic fields increases. In order to check this fact we plot
in Figure~\ref{fig:condensate}~(left) the dependence of lowest Dirac eigenvalues on the strength of the magnetic field.
One can clearly see that the increase of the strength of the magnetic field leads to the consolidation of the near-zero modes.
This configuration has topological charge equal to one, and therefore the configuration contains one exact zero mode, $\lambda=0$,
in agreement with the Atyah-Singer theorem. The presence of the zero mode is independent of the strength of the external Abelian magnetic field
because the number of zero eigenmodes is equal to the topological charge of the gauge field configuration,
while the topological charge is not affected by the Abelian magnetic field.

In Figure~\ref{fig:condensate}~(right) we show the chiral condensate as the function of the magnetic field $q B$
at zero temperature and at $T=0.82\, T_c$. Following \eq{cc_vs_B_chPT} we fit the chiral condensate by the linear function,
\begin{eqnarray}
\label{fit_fun}
\Sigma\lr{B} = \Sigma_0\,\lr{1 + \frac{e B}{\Lambda_B^2}}\,,
\end{eqnarray}
where $\Sigma_0$ and $\Lambda_B$ are the fitting parameters.

The best fits are shown in Figure~\ref{fig:condensate}~(right) by the solid lines.
The best fit parameters at zero and nonzero temperatures are, respectively~\cite{ref:condensate}:
\begin{eqnarray}
\label{fit_best}
T=0: \quad\ \qquad \quad \Sigma^{\mathrm{fit}}_0 & = & {[\lr{320 \pm 5} \, \mbox{MeV}]}^3\,,
\qquad
\Lambda_B^{\mathrm{fit}} = (1.53 \pm 0.11) \, \mbox{GeV},\quad
\\
T=0.82 \, T_c: \quad  \quad \Sigma^{\mathrm{fit}}_0 & = & {[\lr{291 \pm 1} \, \mbox{MeV}]}^3\,,
\qquad
\Lambda_B^{\mathrm{fit}} = (1.74 \pm 0.03) \, \mbox{GeV}.
\quad
\end{eqnarray}
We see that the increase of temperature leads to a decrease the chiral condensate (as expected), and to a decrease the
slope of the $B$-dependence.

The zero-field zero-temperature best fit value~(\ref{fit_best}) of the chiral condensate, $\Sigma^{\mathrm{fit}}_0$, agrees
very well with other numerical estimations in quenched $SU(2)$ gauge theory~\cite{Hands:1990wc}. Surprisingly, the numerical value
of the slope parameter $\Lambda_B$ is quite close to the $T=0$ result of the chiral perturbation theory~(\ref{cc_vs_B_chPT}),
Ref.~\cite{Smilga:97:1}:
\begin{eqnarray}
\label{c:theory}
\Lambda_B^{\mathrm{theory}} = 4 \pi F_{\pi}/\sqrt{\ln{2}} = 1.97 \, \mbox{GeV}\,.
\end{eqnarray}
The similarity between the theoretical prediction~\eq{c:theory} and first principle $T=0$ result~\eq{fit_best} should be taken with care.
First, we are studying the quenched theory, in which the virtual charged pions are absent, and what we observe is the effect of the gauge
fields. Second, the strength of our weakest magnetic fields is still greater than the scale imposed by the zero-field chiral condensate,
$\Sigma^{1/3}_0$. In this regime the prediction~(\ref{cc_vs_B_chPT}) of Ref.~\cite{Smilga:97:1} should not work, in general, as the
linear behavior is expected to be realized for much weaker fields. Thus, the observed linear enhancement of the chiral condensate in
the absence of the pion loops is an intriguing unexpected feature of the non-Abelian gauge theory exposed to very intense external magnetic fields.

\clearpage
\section{Magnetization of QCD vacuum}
\label{sec:magnetization}

Quarks are spin-1/2 particles carrying magnetic moments. The magnetic moments are polarized by the external magnetic field.
A natural quantitative measure of the polarization is given by the expectation
value\footnote{Flavor and spinor indices are omitted in \eq{eq:chi:def}. Below we consider one quark flavor for simplicity.}~\cite{ref:IS}
\begin{eqnarray}
\langle \bar\Psi \Sigma_{\alpha\beta} \Psi\rangle = \chi(F)\, \langle \bar\Psi
\Psi\rangle\, q F_{\alpha\beta}\,, \label{eq:chi:def}
\end{eqnarray}
where $\Sigma_{\alpha\beta} = \frac{1}{2 i} [\gamma_\alpha \gamma_\beta - \gamma_\beta \gamma_\alpha]$
is the relativistic spin operator, $F_{\mu\nu} = \partial_\mu a_\nu -
\partial_\nu a_\mu$ is the strength tensor of the electromagnetic field $a_\mu$.

The right hand side of Eq.~\eq{eq:chi:def} is proportional to the
electromagnetic field strength tensor due to the Lorenz covariance. The
proportionality to the chiral condensate
$\langle \bar\Psi \Psi\rangle$ (evaluated at the nonzero external electromagnetic field~$F$)
allows us to disentangle nonlinear effects of the enhancement
of the chiral condensate in the external magnetic field (discussed in the previous Section)
from the effects of the quark's spin polarization. The strength of the vacuum polarization
is characterized by the chiral magnetic susceptibility $\chi(F)$.

The (chiral) magnetization of the QCD vacuum in the external magnetic
field $B = F_{12} = - F_{21}$ can be described by the dimensionless quantity
\begin{eqnarray}
\mu(q B) = \chi(q B)\, q B
\qquad
\Longleftrightarrow
\qquad
\langle \bar\Psi \Sigma_{12} \Psi\rangle = \mu(q B) \langle \bar\Psi
\Psi\rangle\,.
\label{eq:Sigma12}
\label{eq:mu}
\end{eqnarray}

In \cite{ref:magnetization} we derived a magnetization analogue of the Banks-Casher formula~\eq{BanksCasher}:
\begin{eqnarray}
\langle \bar\Psi \Sigma_{\alpha\beta} \Psi\rangle  = \lim_{\lambda \to 0}
\Bigl\langle\frac{\pi \nu(\lambda)}{V} \int d^4 x \, \psi^\dagger_\lambda(x) \,
\Sigma_{\alpha\beta}\, \psi_\lambda(x)\Bigr\rangle\,. \label{eq:lim4}
\end{eqnarray}
Moreover, we have proven theoretically in \cite{ref:magnetization}
and also checked numerically, that the following factorization rule holds:
\begin{eqnarray}
\langle \bar\Psi \, \Sigma_{\alpha\beta} \, \Psi\rangle = \bigl\langle\bar\Psi
\Psi\bigr\rangle \Bigl\langle\int d^4 x \, \psi^\dagger_\lambda(x) \,
\Sigma_{\alpha\beta} \, \psi_\lambda(x)\Bigr\rangle\,.
\label{eq:lim5}
\label{eq:factorization}
\end{eqnarray}
The comparison of this formula with Eqs.~\eq{eq:chi:def} and \eq{eq:mu} gives
\begin{eqnarray}
\mu(q B) \equiv \chi(q F) \, q F_{\alpha\beta} = \lim_{\lambda \to 0} \Bigl\langle\int d^4 x \, \psi^\dagger_\lambda(x;B)
\, \Sigma_{12} \, \psi_\lambda(x;B)\Bigr\rangle\,,
\label{eq:mu:fin}
\label{eq:chi}
\end{eqnarray}
where $\psi_\lambda(x;F)$ is the eigenmode of the Dirac operator in the external (magnetic) background field $B=F_{12}$.

\begin{figure}[htb]
\begin{center}
\begin{tabular}{cc}
\hskip -7mm \includegraphics[height=3.1in, angle=-90]{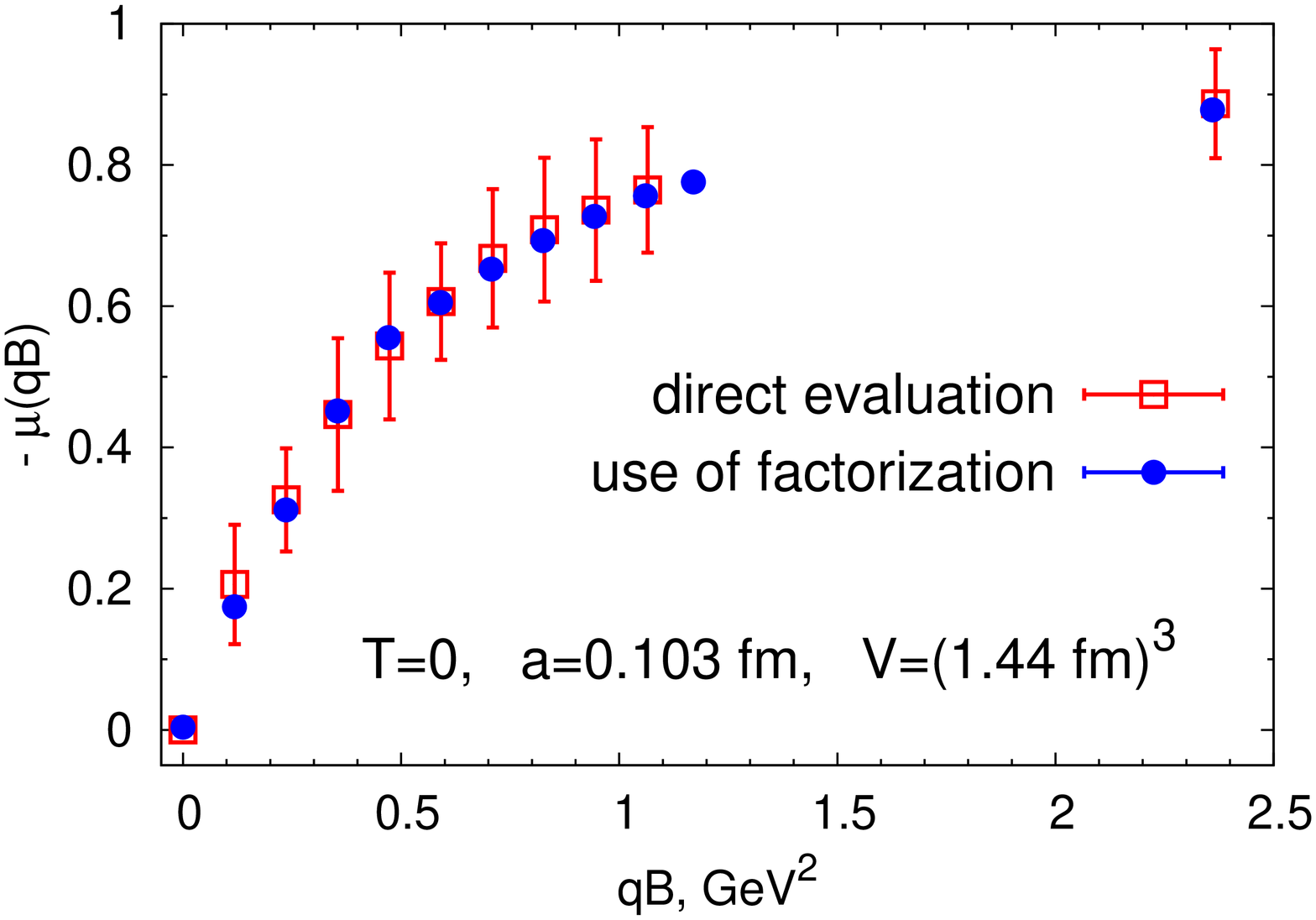} &
\hskip -5mm \includegraphics[height=3.1in, angle=-90]{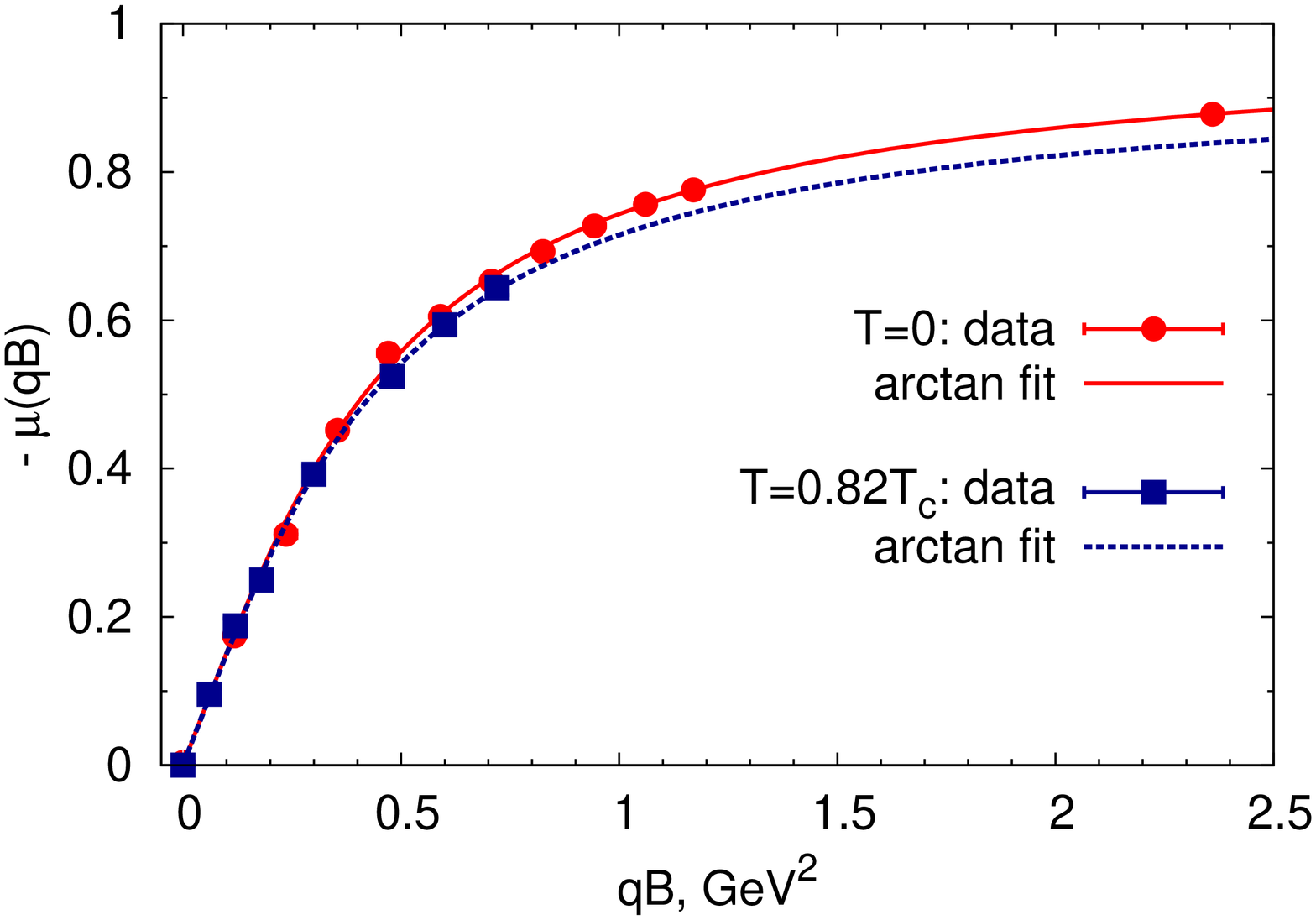}
\end{tabular}
\caption{The magnetization $\mu$ vs the magnetic field $q B$. (left) Check of the factorization~\eq{eq:factorization}:
the empty squares and the full circles show the nonfactorized~\eq{eq:lim4} and factorized~\eq{eq:mu:fin}
definitions, respectively. (right) The magnetization $\mu$ at two temperatures.
The lines correspond to the best fits~\eq{eq:arctan2}. Figures are from Ref.~\cite{ref:magnetization}.}
\label{fig:magnetization}
\end{center}
\end{figure}

We established the factorization property~\eq{eq:lim5} for all studied values of the magnetic fields.
The comparison of the left and right sides of Eq.~\eq{eq:lim5} evaluated at the same set of configurations
is shown in Figure~\ref{fig:magnetization}~(left). While both estimations agree with each other within error bars,
the factorized definition~\eq{eq:mu:fin} is much more accurate compared to the nonfactorized one~\eq{eq:lim4}.

We show in Figure~\ref{fig:magnetization}~(right) the magnetization for two different
temperatures, $T=0$ and $T=0.82 \, T_c$. The behavior of the magnetization is consistent
with general expectations: at low magnetic fields the magnetization is a linear function,
which indicates the existence of a nonzero susceptibility at vanishingly small external magnetic field. At high
magnetic fields the quarks are fully polarized and the
magnetization approaches the saturation regime, $\mu(q B) \to - 1$.

We fit our data by the function (other fitting functions are discussed in \cite{ref:magnetization})
\begin{eqnarray}
\mu^{\mathrm{trig}}_{\mathrm{fit}}(B) & = & \frac{2 \mu_\infty}{\pi}  \arctan \frac{\pi \chi_0 q B}{2 \mu_\infty}\,,
\label{eq:arctan2}
\end{eqnarray}
which has two fitting parameters: the zero-field susceptibility $\chi_0$ and
the infinite-field saturation parameter $\mu_\infty$.
The fits -- shown by the lines in Figure~\ref{fig:magnetization}~(right) --
give the following values for the chiral susceptibility:
\begin{eqnarray}
\chi_0 =
\left\{
\begin{array}{llll}
- 1.547(6) \!\!& \mbox{GeV}^{-2} & \qquad\Lambda_{\mathrm{UV}} \sim 2\,\mbox{GeV}   & \qquad T=0 \\
- 1.53(3)  \!\!& \mbox{GeV}^{-2} & \qquad\Lambda_{\mathrm{UV}} \sim 1.5\,\mbox{GeV} & \qquad T=0.82 T_c
\end{array}
\right.
\label{eq:summary}
\end{eqnarray}
where $T$ is the temperature, and $\Lambda_{\mathrm{UV}}$ is the momentum scale, at which the susceptibility is evaluated
(the momentum dependence is discussed, e.g., in~\cite{Dorokhov:alone}).

An experimentally relevant and phenomenologically interesting quantity is the product $\chi \langle \bar\Psi\Psi\rangle$
at vanishing field~\cite{ref:phenomenology}. Our data give the following result:
\begin{eqnarray}
\chi \, \langle \bar\Psi\Psi\rangle = - 46(3)\,\mbox{MeV}\ \qquad \mbox{[quenched limit]}\,.
\label{eq:chi:cond}
\end{eqnarray}
which is surprisingly close to the estimation based on the QCD sum rules techniques,
$\chi \, \langle \bar\Psi\Psi\rangle \approx - 50\,\mbox{MeV}$~\cite{ref:others:OPE}.

\clearpage
\section{Chiral Magnetic Effect}
\label{sec:CME}

In brief, the physical mechanism behind the Chiral Magnetic Effect (CME) is as follows~\cite{Kharzeev:08:1}.
A strong magnetic field forces the magnetic moment of quarks to turn parallel to the direction of the field.
If the quarks are light (massless) then the left-handed quarks will move, say, towards the direction of the magnetic field while the
right-handed quarks will move backwards. If there is an imbalance between left-handed and right-handed quarks, then a net electric
current $j_\mu\lr{x} = \bar{\psi}(x) \gamma_\mu \psi(x)$ appears along the magnetic
field~\cite{Kharzeev:08:1,Kharzeev:98:1,Nielsen:83:1}.
This longitudinal electric current
may lead also to a spatial separation of the electric charges. The longitudinal current and the spatial charge separation are
the consequences of the CME. The chiral imbalance may be created by topologically nontrivial configurations of gluon fields.
Generally, the CME is a reflection of the $\cal{CP}$-odd structure of the vacuum. The existence of the CME was demonstrated
in lattice simulations in Ref.~\cite{ref:CME}.

\begin{figure}[htb]
\begin{center}
\begin{tabular}{cc}
& \\[-15mm]
\includegraphics[height=2.0in, angle=0]{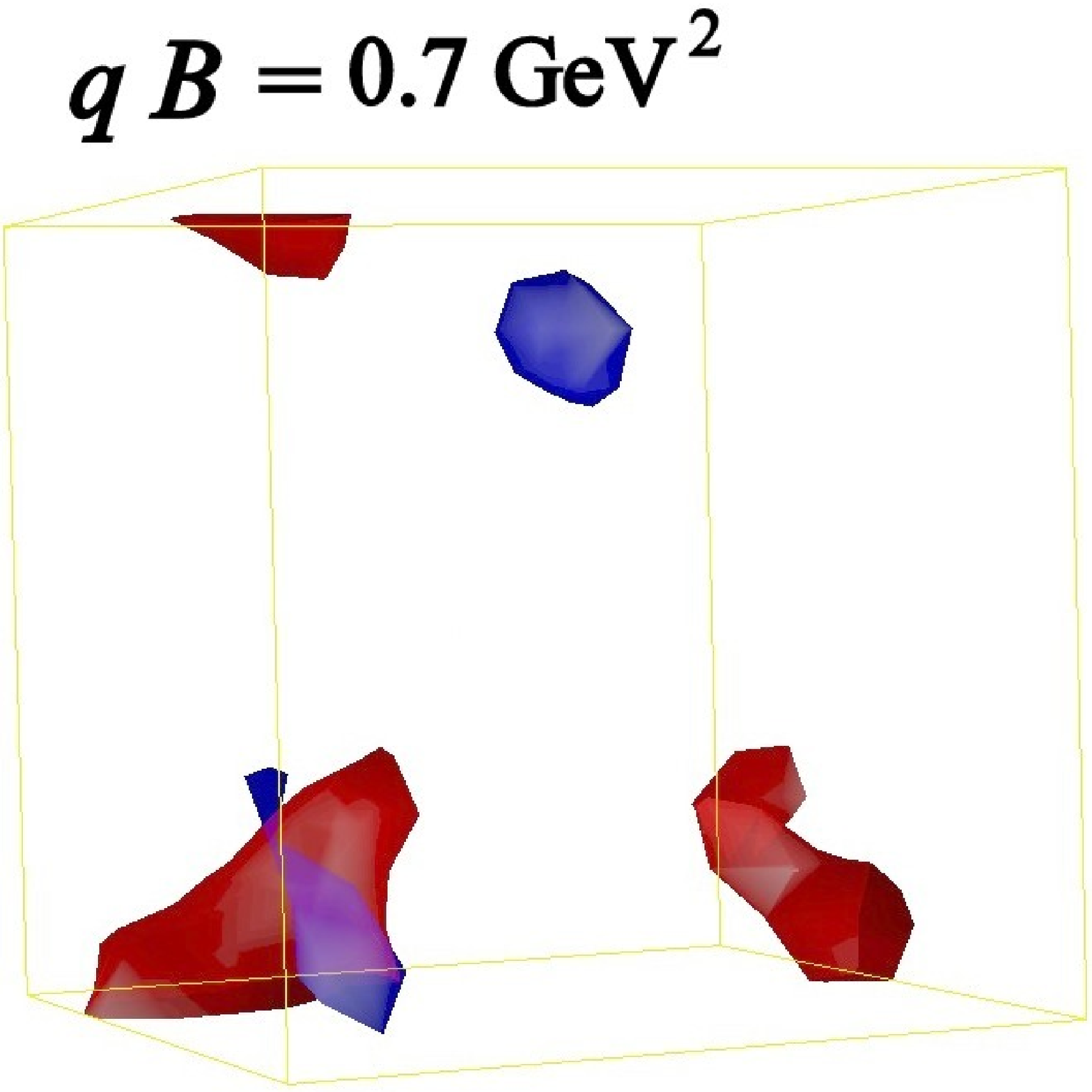} &
\includegraphics[height=2.0in, angle=0]{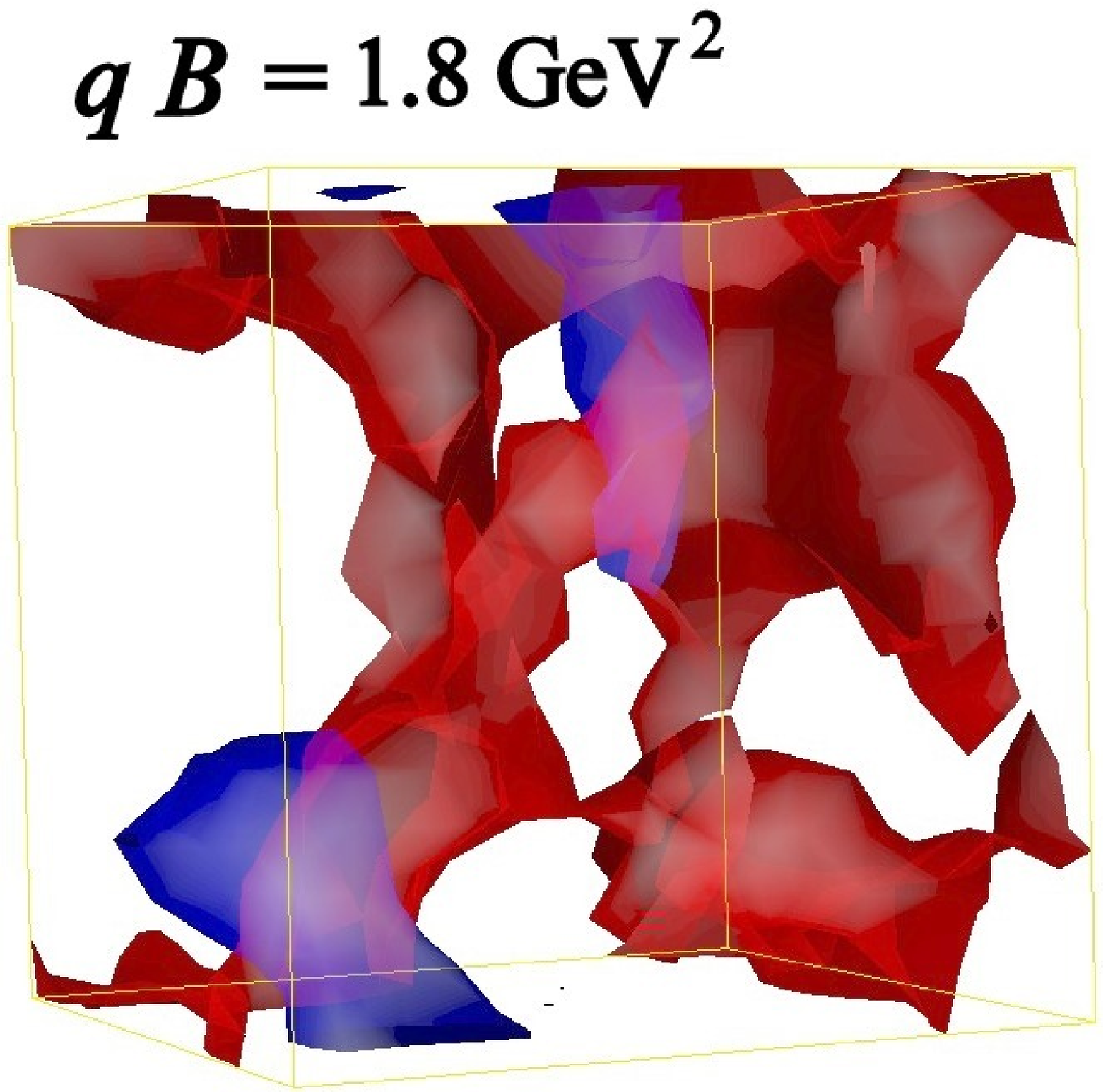}
\end{tabular}
\caption{A visual evidence of the CME at a typical $T=0$ gluon configuration: the excess of the positive (red) and negative (blue)
electric charge due to the magnetic field (directed vertically) $qB = 0.7\,\mbox{GeV}^2$ (left) and $qB = 1.8\,\mbox{GeV}^2$ (right), Ref.~\cite{ref:CME}.}
\label{fig:CME:charge}
\end{center}
\end{figure}

In Figure~\ref{fig:CME:charge} we plot a spatial excess of the charge density originating due to the applied magnetic field,
$j_0\lr{x;B} = \vev{j_0\lr{x}}_{B} - \vev{j_0\lr{x}}_{B=0}$. The subtraction of the zero-field density removes all
ultraviolet contributions providing us with a nonperturbative quantity.
The magnetic field applied to a typical gluon configuration leads
to a spatial separation of the electric charges in an agreement with the CME.
The effect increases as the magnetic field gets stronger.

\begin{figure}[htb]
\begin{center}
\begin{tabular}{cc}
& \\[-12mm]
\includegraphics[height=1.7in, angle=0]{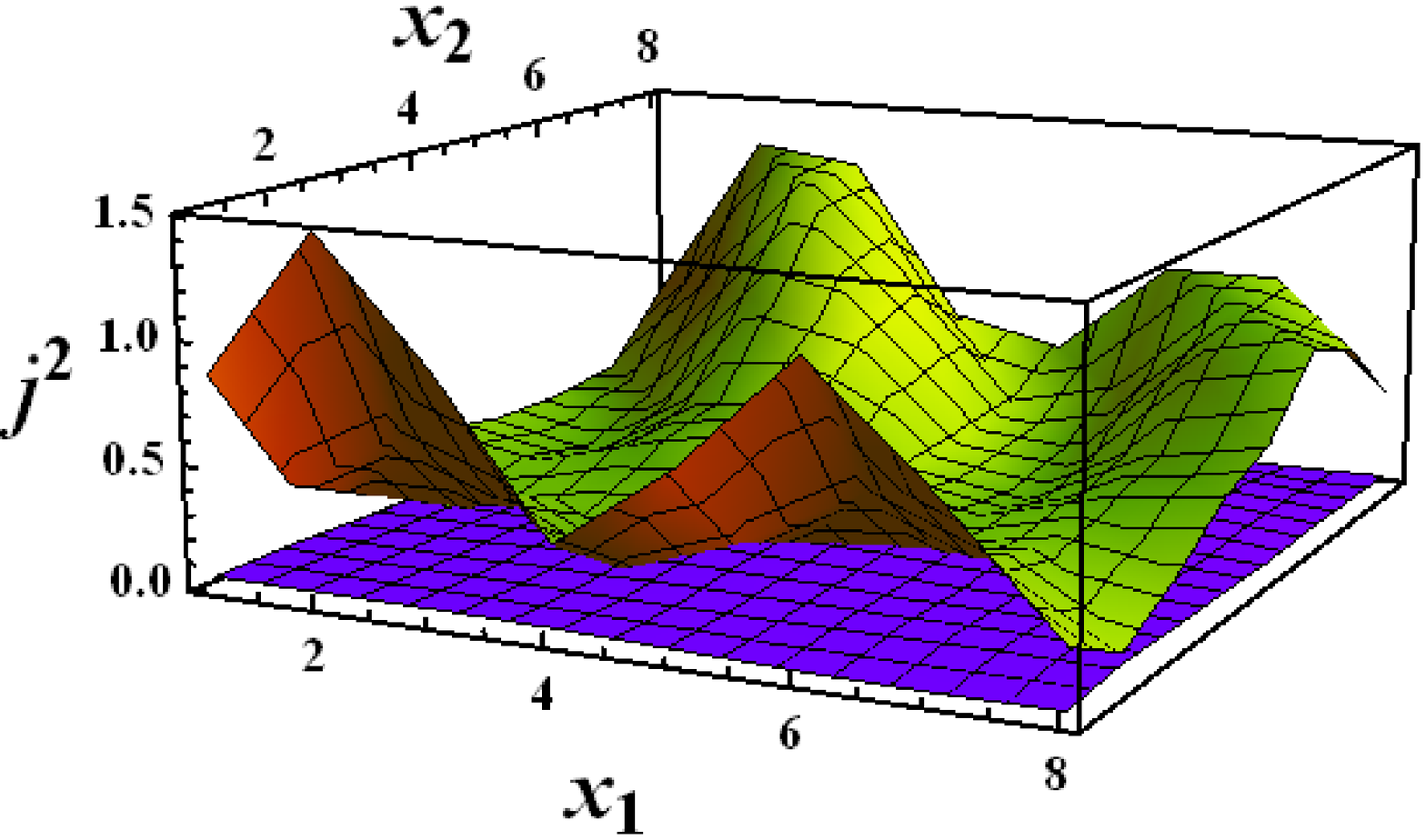} &
\includegraphics[height=1.7in, angle=0]{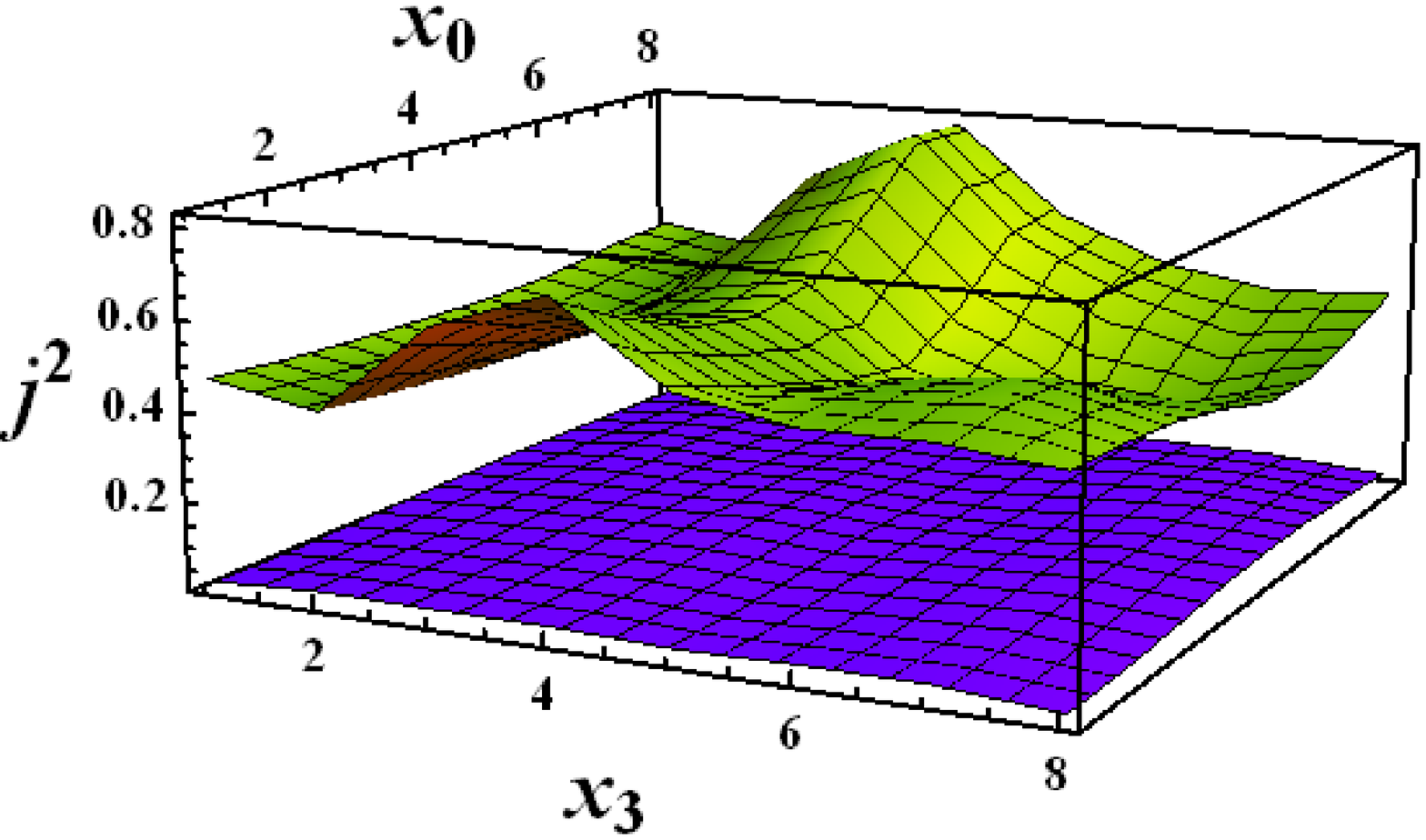}
\end{tabular}
\caption{An evidence of the CME at an instanton-like configuration of unit topolo\-gi\-cal charge:
a magnetic field induces the electric current $j$. The upper (green) and lower (blue) surfaces are the
longitudinal ($j^2_\parallel = j_0^2 + j_3^2$) and the transverse ($j^2_\perp = j_1^2 + j_2^2$)
currents in the $12$-plane (left) and in the $30$-plane (right), Ref.~\cite{ref:CME}.}
\label{fig:CME:inst}
\end{center}
\end{figure}

In Figure~\ref{fig:CME:inst} the profile of the longitudinal (parallel to the magnetic field) and transverse (perpendicular to the field) components
of the quark's electric current in a background of a specially-prepared smooth instanton-like ($Q=1$) configuration
subjected to an external magnetic field. The CME is characterized by enhancement of the longitudinal current with
respect to the transverse one. This property is clearly seen in the case of the instanton, Figure~\ref{fig:CME:inst}.

\begin{figure}[htb]
\begin{center}
\begin{tabular}{cc}
& \\[-12mm]
\hskip -7mm \includegraphics[height=2.7in, angle=-90]{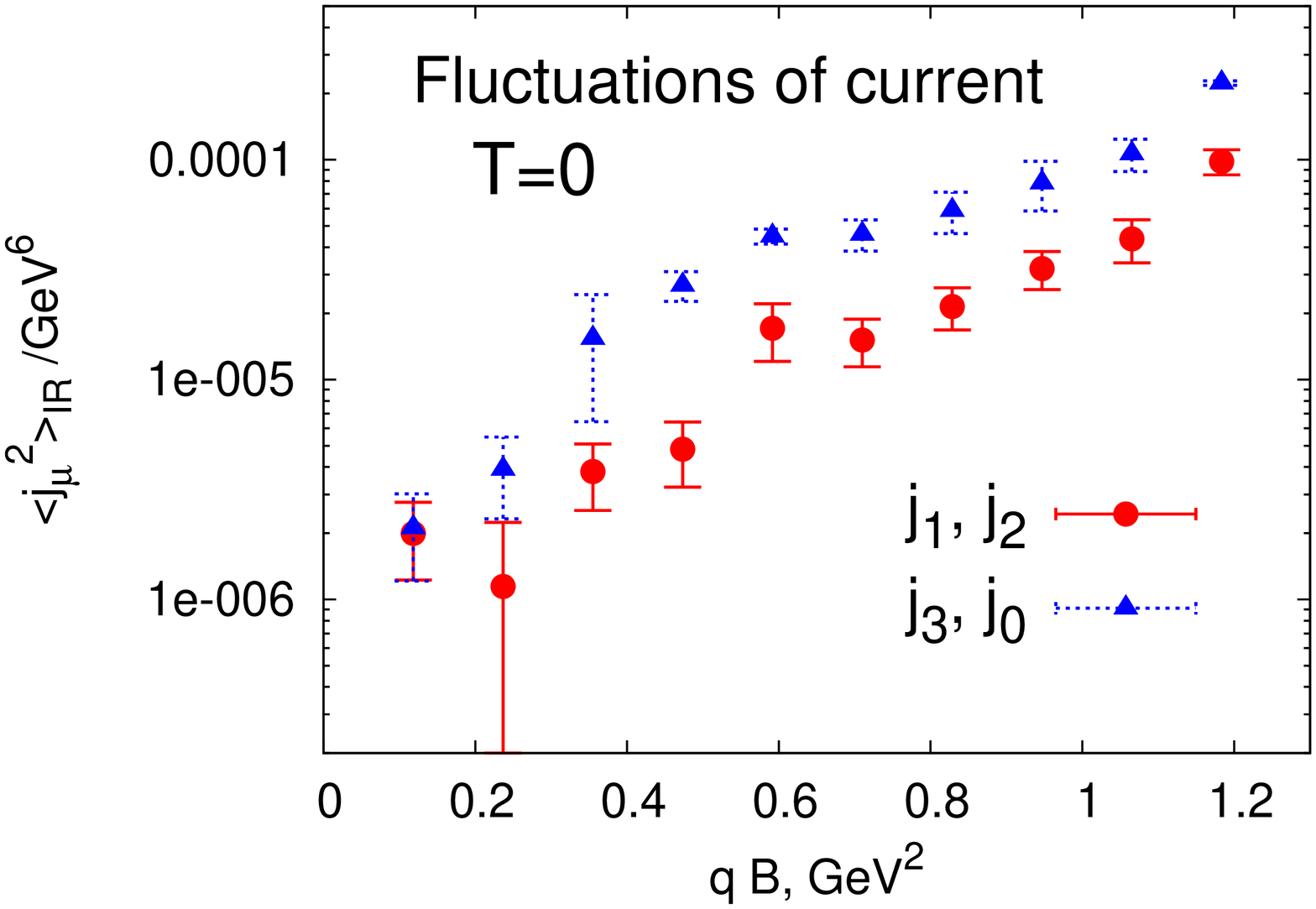} &
\hskip -5mm \includegraphics[height=2.7in, angle=-90]{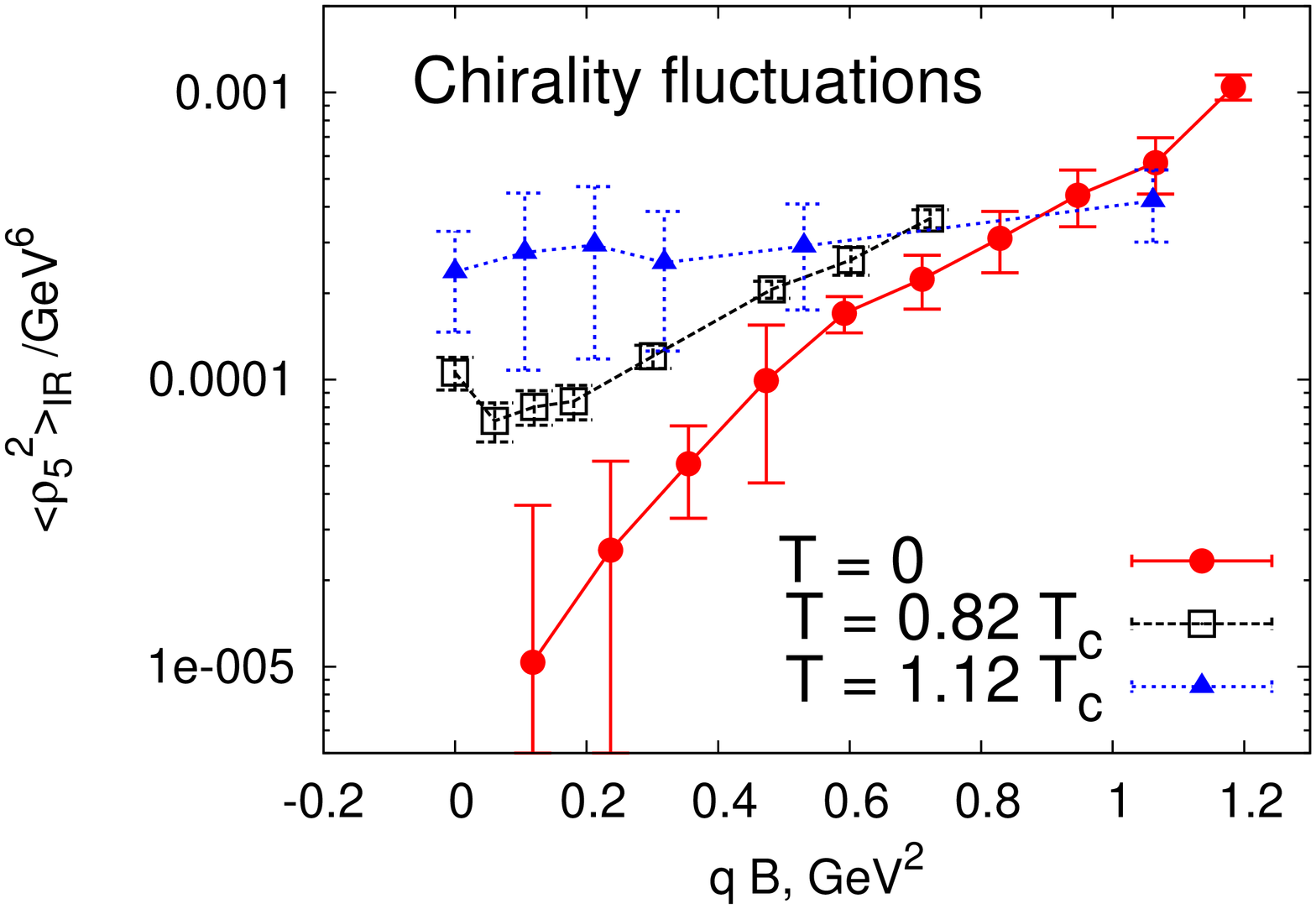}
\end{tabular}
\caption{(left) The squares of the transverse ($j_1$, $j_2$) and longitudinal ($j_3$, $j_0$)
components of the current vs the magnetic field $qB$ at $T=0$.
(right) The chirality squared vs the magnetic field $qB$ at various temperatures $T$. Figures are from Ref.~\cite{ref:CME}.}
\label{fig:CME:rho5}
\end{center}
\end{figure}

Coming back to real quantum configurations of the gluonic fields, we plot
in Figure~\ref{fig:CME:rho5}~(left) the expectation values of the fluctuations
of each component of the electric current at zero temperature. We do not distinguish
the global topological charge of each configuration since the CME effect appears
also locally (an extended discussion of this point is given in Ref.~\cite{ref:CME}).
In agreement with the CME features, we observe the dominance of the longitudinal components
of the electric current with respect to the transverse ones.

The CME crucially depends on the (local) imbalance between the left and right chiral modes given by the local chiral charge
$\rho_{5}\lr{x} = \bar{\psi}\lr{x} \gamma_{5} \psi\lr{x}$. The local strength of the chiral imbalance
is characterized the expectation value of the chirality fluctuations, $\langle \rho_5^2 \rangle$, which is shown
in Figure~\ref{fig:CME:rho5}~(right) for three different temperatures.
At $T=0$ the fluctuations quickly grow with the increase of the magnetic field. As the temperature increases the growth rate gets smaller,
and in the deconfinement phase the rate is almost zero.
In the confinement phase ($T=0$) the chirality fluctuations at strong enough magnetic fields
are as large as the fluctuations in the deconfinement phase. This fact allows us to suggest that
the CME may be observed in a cold nuclear matter as well.


\clearpage
\section{Quark electric dipole moment}
\label{sec:EDM}

The CME induces the electric dipole moment due to the spatial charge separation along the direction of the magnetic field.
Another manifestation of a nontrivial topological structure of the QCD vacuum is the appearance of the {\it quark} electric dipole
moment directed along the axis of the external magnetic field. As in the case of the CME, the bulk average of
the quark electric dipole moment is zero ($\left\langle \bar\Psi {\vec \sigma}^E \Psi \right\rangle = 0$)
due to the global CP-invariance of the QCD vacuum.
However, this anomalous effect may be seen on the event-by-event basis: locally the electric dipole moment of a quark may be large,
while its sign may be alternating as the quark travels trough local fluctuations of the topological charge.
In order to evaluate the magnitude of the local electric dipole moment of the quark, we study the following connected expectation values:
\begin{eqnarray}
{\bigl\langle \bigl(\sigma^\ell_3\bigr)^2 \bigr\rangle}_{IR}
& = & {\Bigl\langle\bigl(\sigma^\ell_3 - \bigl\langle\sigma^\ell_3\bigr\rangle\bigr)^2\Bigr\rangle}_{B, T}
- {\Bigl\langle \bigl(\sigma^\ell_3 - \bigl\langle\sigma^\ell_3\bigr\rangle\bigr)^2\Bigr\rangle}_{B,T = 0}\,, \quad \ell = E,\, M\,, \quad
\label{ref:expectation}
\end{eqnarray}
where $\sigma^E_i(x) = \bar{\psi}\lr{x} \Sigma_{i0} \psi\lr{x}$ is the local density of the electric dipole moment [the density of its magnetic counterpart is $\sigma^M_i(x) = \frac{1}{2} \varepsilon_{ijk} \bar{\psi}\lr{x} \Sigma_{jk} \psi\lr{x}$].
\begin{figure}[htb]
\begin{center}
\vskip -5mm
\begin{tabular}{cc}
\hskip -7mm \includegraphics[height=3.1in, angle=-90]{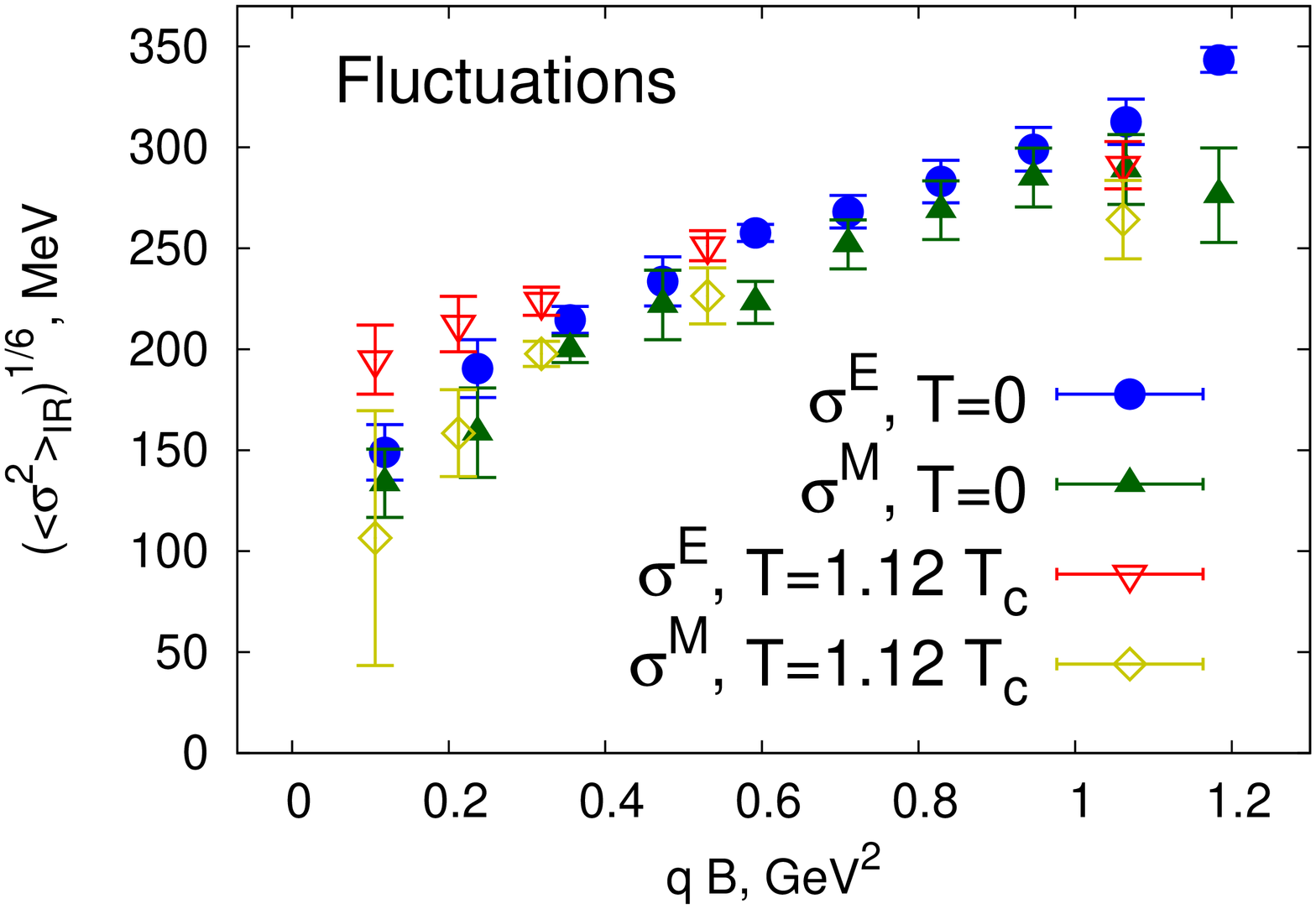} &
\hskip -5mm \includegraphics[height=3.1in, angle=-90]{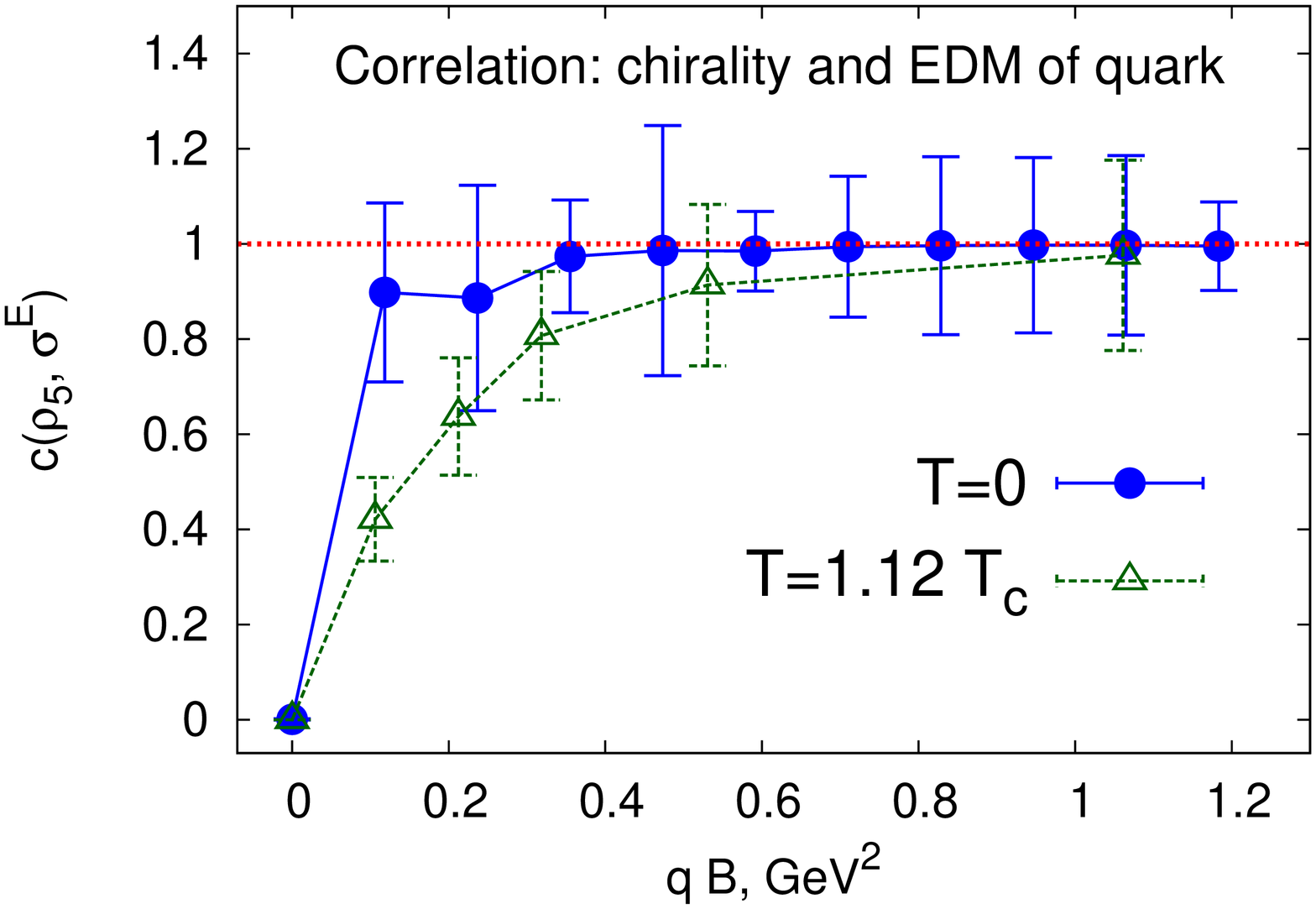}
\end{tabular}
\caption{(left) The fluctuations of the longitudinal components the magnetic and electric dipole densities vs $qB$ at $T=0$ and $T = 1.12 \, T_c$.
(right) The correlation of the electric dipole moment of quark (EDM) with the chiral density vs $qB$.}
\vskip -5mm
\label{fig:spin}
\end{center}
\end{figure}

In Figure~\ref{fig:spin}~(left) we show values of the fluctuations of the longitudinal (i.e.,
directed along the magnetic field) component of the dipole moments~\eq{ref:expectation}.
The fluctuations of the electric and magnetic dipole moments
are equally strong in both phases. The transverse fluctuations of both electric and magnetic dipole moments are zero within our error bars.

The electric dipole moment of the quark is closely related to the local chirality. In Figure~\ref{fig:spin}~(right) we plot the (normalized)
correlator of the electric dipole moment with the chiral density,
$c\lr{ \rho_{5}, \sigma^E_3 } = \langle \rho_{5} \, \sigma^E_3 \rangle/[{\langle \rho_{5}^{2}\rangle} {\langle(\sigma^E_3)^{2}\rangle}]^{1/2}$.
At strong magnetic field we observe the very strong correlation of these two quantities in both confinement and deconfinement phases,
while at weaker magnetic fields the correlation decreases. We also observed that the thermal fluctuations
suppress the correlation function $c\lr{ \rho_{5}, \sigma^E_3 }$, and that magnetic moment is not correlated with the local chirality.

Thus, we found an evidence that the external magnetic field forces the quark to develop a local electric dipole moment
along the direction of the field.

\section*{Acknowledgments}
The authors are grateful to Ph.~Boucaud, V.G.~Bornyakov, V.V.~Bra\-gu\-ta,
A.S.~Gorsky, B.L.~Ioffe, B.O.~Kerbikov, D.~Kharzeev, A.~Krikun, S.M.~Morozov,
V.A.~Novikov, B.~Pire, V.I.~Shevchenko, M.I.~Vysotsky, and V.I.~Zakharov for
interesting discussions and suggestions.
This work was partly supported by Grants RFBR Nos. 08-02-00661-a and 09-02-00629, grants for scientific schools
Nos. NSh-679.2008.2 and Nsh-4961.2008.2, by the Russian Federal  Agency for Nuclear Power, and by the STINT
Institutional grant IG2004-2 025. P.V.B. is also partially supported by the Euler scholarship
from DAAD, by a scholarship of the Dynasty Foundation and by the grant BRFBR
F08D-005 of the Belarusian Foundation for Fundamental Research. The calculations were partially done
on the MVS 50K at Moscow Joint Supercomputer Center.


\begin{thebibliography}{99}

\bibitem{Kharzeev:08:1}
  D.E.~Kharzeev, L.D.~McLerran, and H.J.~Warringa,
  Nucl.\ Phys.\  A {\bf 803}, 227 (2008) [arXiv:0711.0950];
 K.~Fukushima, D.E.~Kharzeev, and H.J.~Warringa,
  Phys.\ Rev.\  D {\bf 78}, 074033 (2008) [arXiv:0808.3382];
 H.J.~Warringa,
  arXiv:0906.2803;
 D.E.~Kharzeev,
  arXiv:0906.2808 and
  arXiv:0908.0314.


\bibitem{Skokov}
V.~Skokov, A.~Illarionov and V.~Toneev,
  arXiv:0907.1396.

\bibitem{Agasian}
  N.~O.~Agasian and S.~M.~Fedorov,
  Phys.\ Lett.\  B {\bf 663}, 445 (2008).

\bibitem{Fraga:08:1:2:3}
  E.~S.~Fraga and A.~J.~Mizher,
  Phys.\ Rev.\  D {\bf 78}, 025016 (2008) [arXiv:0804.1452];
Nucl.\ Phys.\  A {\bf 820}, 103C (2009) [arXiv:0810.3693].

\bibitem{Smilga:97:1}
  I.~A.~Shushpanov and A.~V.~Smilga,
  Phys.\ Lett.\  B {\bf 402}, 351 (1997).

\bibitem{Agasian:1999sx}
  N.~O.~Agasian and I.~A.~Shushpanov,
  Phys.\ Lett.\  B {\bf 472}, 143 (2000).

\bibitem{Cohen:2007bt}
  T.D. Cohen, D.A. McGady, E.S. Werbos,
  Phys.\ Rev.\  C {\bf 76}, 055201 (2007).

\bibitem{Gusynin:1995nb}
  S.~P.~Klevansky and R.~H.~Lemmer,
  Phys.\ Rev.\  D {\bf 39}, 3478 (1989);
  V.P.Gusynin, V.A.Miransky, I.A.Shovkovy,
  Nucl.\ Phys.\  B {\bf 462}, 249 (1996) [hep-ph/9509320];
  D.~Ebert, K.~G.~Klimenko, M.~A.~Vdovichenko and A.~S.~Vshivtsev,
  Phys.\ Rev.\  D {\bf 61}, 025005 (1999) [hep-ph/9905253].

\bibitem{Goyal:1999ye}
  A.~Goyal, M.~Dahiya,
  Phys.\ Rev.\  D {\bf 62}, 025022 (2000) [hep-ph/9906367].

\bibitem{Zayakin:08:1}
  A.~V.~Zayakin,
  JHEP {\bf 0807}, 116 (2008) [arXiv:0807.2917].

\bibitem{ref:condensate}
  P.V.~Buividovich, M.N.~Chernodub, E.V.~Luschevskaya, M.I.~Polikarpov,
  {\sl ``Numerical study of chiral symmetry breaking ...''}, arXiv:0812.1740.

\bibitem{ref:IS}
B.~L.~Ioffe and A.~V.~Smilga,
  Nucl.\ Phys.\  B {\bf 232}, 109 (1984).

\bibitem{ref:others:OPE}
  I.~I.~Balitsky, A.~V.~Yung,
  Phys.\ Lett.\  B {\bf 129}, 328 (1983);
V.~M.~Belyaev, Y.~I.~Kogan,
  Yad.\ Fiz.\  {\bf 40}, 1035 (1984);
  I.~I.~Balitsky, A.~V.~Kolesnichenko, A.~V.~Yung,
  Sov.\ J.\ Nucl.\ Phys.\  {\bf 41}, 178 (1985);
  P.~Ball, V.~M.~Braun, N.~Kivel,
  Nucl.\ Phys.\  B {\bf 649}, 263 (2003) [arXiv:hep-ph/0207307].

\bibitem{Dorokhov:alone}
A.~E.~Dorokhov,
  Eur.\ Phys.\ J.\  C {\bf 42}, 309 (2005)
  [arXiv:hep-ph/0505007].

\bibitem{Dorokhov}
A.~E.~Dorokhov, W.~Broniowski and E.~Ruiz Arriola,
  Phys.\ Rev.\  D {\bf 74}, 054023 (2006)
  [arXiv:hep-ph/0607171];
  arXiv:0907.3374 [hep-ph].

\bibitem{ref:Vainshtein}
  A.~Vainshtein,
  Phys.\ Lett.\  B {\bf 569}, 187 (2003)
  [arXiv:hep-ph/0212231].

\bibitem{ref:Sasha}
  A.~Gorsky and A.~Krikun,
  arXiv:0902.1832 [hep-ph].

\bibitem{ref:Kim}
  H.~C.~Kim, M.~Musakhanov, M.~Siddikov,
  Phys.\ Lett.\  B {\bf 608}, 95 (2005).

\bibitem{ref:phenomenology}
V.~M.~Braun, S.~Gottwald, D.~Y.~Ivanov, A.~Schafer, L.~Szymanowski,
  Phys.\ Rev.\ Lett.\  {\bf 89}, 172001 (2002);
B.~Pire, L.~Szymanowski,
Phys.\ Rev.\ Lett.\  {\bf 103}, 072002 (2009) [arXiv:0905.1258];
arXiv:0909.0098 [hep-ph].

\bibitem{Rohrwild:2007yt}
  J.~Rohrwild,
  JHEP {\bf 0709}, 073 (2007)
  [arXiv:0708.1405 [hep-ph]].

\bibitem{ref:Thomas}
  T.D.Cohen,  E.S.Werbos,
  Phys.Rev. C 80, 015203 (2009) [arXiv:0810.5103].

\bibitem{ref:magnetization}
  P.V.~Buividovich, M.N.~Chernodub, E.V.~Luschevskaya, M.I.~Polikarpov,
  {\sl ``Chiral magnetization of non-Abelian vacuum ...''}, arXiv:0906.0488.

\bibitem{Kharzeev:98:1}
  D.~Kharzeev, R.D.~Pisarski, M. Tytgat,
  Phys. Rev. Lett.\  {\bf 81}, 512 (1998) [hep-ph/9804221];
  D.~Kharzeev,
  Phys.\ Lett.\  B {\bf 633}, 260 (2006).

\bibitem{Voloshin:04:1}
  S.~A.~Voloshin,
  Phys.\ Rev.\  C {\bf 70}, 057901 (2004)
  [arXiv:hep-ph/0406311].

\bibitem{Voloshin:08:1}
  S.~A.~Voloshin  [STAR Collaboration],
  arXiv:0806.0029 [nucl-ex];
H.~Caines [STAR Collaboration], arXiv:0906.0305 [nucl-ex].

\bibitem{ref:CME}
  P.V.~Buividovich, M.N.~Chernodub, E.V.~Luschevskaya, M.I.~Polikarpov,
  {\sl ``CME in LGT''}, Pis'ma v ZhETF {\bf 90}, 456 (2009) and arXiv:0907.0494.

\bibitem{ref:TBA}
  P.V.~Buividovich, M.N.~Chernodub, E.V.~Luschevskaya, M.I.~Po\-li\-kar\-pov,
  {\sl ``Quark electric dipole moment induced by magn. field''}, arXiv:0909.2350.

\bibitem{Luschevskaya:08:1}
  V.~G.~Bornyakov et al,
  Phys.\ Rev.\  D {\bf 79}, 054505 (2009).

\bibitem{Neuberger:98:1}
  H.~Neuberger,
  Phys.\ Lett.\  B {\bf 417}, 141 (1998)
  [arXiv:hep-lat/9707022].

\bibitem{Wiese:08:1}
  M.~H.~Al-Hashimi and U.~J.~Wiese,
  Annals Phys.\  {\bf 324}, 343 (2009).

\bibitem{Damgaard:88:1}
  P.~H.~Damgaard and U.~M.~Heller,
  Nucl.\ Phys.\  B {\bf 309}, 625 (1988).

\bibitem{Banks:80:1}
  T.~Banks and A.~Casher,
  Nucl.\ Phys.\  B {\bf 169}, 103 (1980).

\bibitem{Hands:1990wc}
  S.~J.~Hands and M.~Teper,
  Nucl.\ Phys.\  B {\bf 347}, 819 (1990).

\bibitem{Nielsen:83:1}
  H.~B.~Nielsen and M.~Ninomiya,
  Phys.\ Lett.\  B {\bf 130}, 389 (1983);
  M.~A.~Metlitski and A.~R.~Zhitnitsky,
  Phys.\ Rev.\  D {\bf 72}, 045011 (2005).

\end{thebibliography}
\end{document}